\documentclass[runningheads]{llncs}
\usepackage{graphicx}
\usepackage{url}
\usepackage[hidelinks]{hyperref}
\usepackage{caption}
\usepackage{amsmath}
\usepackage{xfrac}
\usepackage{xcolor}
\usepackage{bbm}
\usepackage{booktabs}
\usepackage{multirow}
\usepackage{algorithm}
\usepackage[noend]{algorithmic}
\usepackage[labelformat=simple]{subcaption}

\newtheorem{mydef}{Definition}
\begin{document}
\title{Selfish Mining in Proof-of-Work Blockchain with Multiple Miners: An Empirical Evaluation\textcolor{red}{\thanks{\textcolor{red}{This is the author's accepted version of the work. The final authenticated version is maintained by Springer Nature Switzerland AG and is available online at \url{https://doi.org/10.1007/978-3-030-33792-6_14}.}}}}
\titlerunning{Selfish Mining in Proof-of-Work Blockchain with Multiple Miners}
\author{Submission Number 41}
\author{Tin Leelavimolsilp\inst{1}\thanks{Corresponding author} \and Viet Nguyen\inst{2} \and Sebastian Stein\inst{1} \and \\ Long Tran-Thanh\inst{1}}
\authorrunning{T. Leelavimolsilp et al.}
% \institute{}
\institute{
University of Southampton, Southampton, SO17 1BJ, UK \\ \email{\{\href{mailto:tin.leelavimolsilp@soton.ac.uk}{tin.leelavimolsilp},\href{mailto:s.stein@soton.ac.uk}{s.stein},\href{mailto:l.tran-thanh@soton.ac.uk}{l.tran-thanh}\}@soton.ac.uk} \and 
Imperial College London, London, SW7 2AZ, UK \\
\email{\href{mailto:viet.nguyen17@imperial.ac.uk}{viet.nguyen17@imperial.ac.uk}}}
\maketitle
\begin{abstract}
Proof-of-Work blockchain, despite its numerous benefits, is still not an entirely secure technology due to the existence of Selfish Mining (SM) strategies that can disrupt the system and its mining economy. While the effect of SM has been studied mostly in a two-miners scenario, it has not been investigated in a more practical context where there are multiple malicious miners individually performing SM.
To fill this gap, we carry out an empirical study that separately accounts for different numbers of SM miners (who always perform SM) and strategic miners (who choose either SM or Nakamoto's mining protocol depending on which maximises their individual mining reward). 
Our result shows that SM is generally more effective as the number of SM miners increases, however its effectiveness does not vary in the presence of a large number of strategic miners. Under specific mining power distributions, we also demonstrate that multiple miners can perform SM and simultaneously gain higher mining rewards than they should. Surprisingly, we also show that the more strategic miners there are, the more robust the systems become. Since blockchain miners should naturally be seen as self-interested strategic miners, our findings encourage blockchain system developers and engineers to attract as many miners as possible to prevent SM and similar behaviour.
\keywords{Selfish mining \and Proof-of-Work blockchain \and Agent-based model \and Empirical multiplayer game}
\end{abstract}
%
%%%%%%%%%%%%%%%%%%%%%%%%%%%%%%%%%%%
\section{Introduction}
\label{sec:introduction}
%%%%%%%%%%%%%%%%%%%%%%%%%%%%%%%%%%%
With the aim to decrease reliance on financial institutions, blockchain was designed and used to securely approve and record transactions among Internet users \cite{Nakamoto2008}. A number of blockchain characteristics such as its security, transparency, and decentralised authority have drawn many researchers and developers to apply blockchain to a wide range of application areas, such as personal data management \cite{AzariaEtAl2016,ZyskindEtAl2015}, Internet of Things \cite{ChristidisDevetsikiotis2016}, and decentralised platform as a service \cite{Wood2014}. 

The success of blockchain is based on two elements: an application of a cryptographic puzzle, namely Proof-of-Work (PoW), and an economic incentive for miners, who are the underlying workforce of the system. The mining process is briefly described as follows. First, a miner composes a block which mainly consists of locally verified transactions. The block also refers to the latest block of the miner's locally stored blockchain as its parent block. The miner then performs a brute force search for a number that results in a hash value of the block lower than the globally set target. When such a number (which is a ``Proof of Work'' that the miner did) is found, the block together with the number is broadcasted. Subsequently, a recipient of the block verifies the block's transactions and the block's hash value. Once approved, the block is then appended to the recipient's locally stored blockchain. Later, the miner claims their mining reward (which is the aforementioned incentive) by referring to the block in their spending transaction. As such, every miner is fairly rewarded in proportion to a number of blocks that they managed to create or an amount of hash rate that they expended.\footnote[1]{To be precise, there are two types of mining reward: namely, block reward and transaction fee \cite{BitcoinWiki_MiningReward_2018}. While there will be no block reward per block in the future, miners will still be incentivised by the transaction fee to do their mining.}

One of the most fundamental and significant attacks against blockchain systems is \textit{forking}, which is difficult in practice and widely known as the 51\% attack. Since the mining protocol instructs everyone to trust the longest chain\footnotemark[2], a malicious miner simply needs to produce a blockchain longer than the current one. Once succeeded, part of the current blockchain will be replaced by the malicious miner's blocks. Consequently, all transactions and the mining reward of the replaced blocks have been nullified, and the malicious miner earns all mining reward from their blocks; thus resulting in a disproportionate reward distribution. However, forking is not easy since it requires at least a half of the total hash rate in the system \cite{Nakamoto2008}. As such, it resulted in a public belief that blockchain systems are strongly secure as long as no miner possesses more than 50\% of the total hash rate.
\footnotetext[2]{In practice, a chain that is the most computationally expensive or has the highest difficulty sum is always chosen \cite{BitcoinWiki_Block_2019}. If every block has the same computational difficulty (as assumed in this work), the actual verification reduces to selecting the longest blockchain.}

Eyal and Sirer later demonstrated that forking is still possible with lower hash rates using their \textit{Selfish Mining} (SM) strategy \cite{EyalSirer2014}. Essentially, SM hides and privately mines their own blocks in contrast to publicly forking the blockchain. Such hiding allows the malicious miner to gain an advantage by removing the chance of the successive blocks being mined by the others. In addition, SM gradually discloses their private blocks to keep the advantage as much as possible to themselves. Most importantly, it requires only \sfrac{1}{3} of the total hash rate to fork the blockchain and earn a higher mining reward than they should. Such a low hash rate is significantly lower than \sfrac{1}{2} of the total hash rate for publicly forking, and therefore greatly threatens the security of blockchain systems. With a larger hash rate, SM is even more effective and can fork the blockchain more frequently. In the worst scenario, the mining economy and the system could collapse due to the disrupted distribution of the economic incentive.

Moreover, SM can be difficult to detect in practice. While the rate of orphaned blocks (i.e. blocks that were not part of the longest chain) is a main indicator of SM activity \cite{GobelEtAl2016}, it can point out a network instability or a high network delay that causes broadcasted blocks to arrive late or be lost. As such, the practicality of the detection method based on the orphaned block rate is not certain.

Despite the threats posed by SM, there are not sufficient investigations in a more practical context: that is, a case where SM being used individually and simultaneously by multiple miners. In particular, most research so far focused on a system with one malicious miner who performs SM and has another who follows Nakamoto's mining protocol \cite{EyalSirer2014,GervaisEtAl2016,KiayiasEtAl2016,NayakEtAl2016,SapirshteinEtAl2017,ZhangPreneel2017}. In practice, multiple miners can perform SM at the same time. Whether SM is even more effective in such situation is not clearly known.

For this reason, we carry out an investigation on SM in the context of multiple miners. Particularly, we seek to know (a) the effectiveness of SM in such a context, (b) the minimum hash rate that SM requires to earn mining reward more than it should, and (c) the minimum hash rate that non-malicious miners require to prevent SM. We also consider strategic miners who choose either SM or Nakamoto's mining protocol depending on which gives a higher mining reward. We believe that such miners better represent the actual miners since earning mining reward is th e main purpose of their mining activity and they would prefer a higher reward.

The rest of this paper is structured as follows. First, a literature review of existing studies on SM is presented. We then describe two models of PoW blockchain systems (where one considers strategic miners and the another does not) and some concepts that are necessary for our work. Subsequently, our empirical results for each model are described and discussed. We finally conclude this paper with our findings and interesting questions that remain to be solved.
%%%%%%%%%%%%%%%%%%%%%%%%%%%%%%%%%%%%%%%%%%%%%%%%%%
\section{Related Work}
\label{sec:related_work}
%%%%%%%%%%%%%%%%%%%%%%%%%%%%%%%%%%%%%%%%%%%%%%%%%%
After Eyal and Sirer's work, there has been further research on improving SM. To exemplify this, the optimised (two-miners) SM strategy was proposed and its effectiveness was slightly improved \cite{GervaisEtAl2016,SapirshteinEtAl2017}. A combination of SM with other attacks was also designed to increase the effectiveness of the attack \cite{NayakEtAl2016}. In general, such improvements further reduce the amount of required hash rate to successfully employ SM.

A number of studies also shed more light on SM under different contexts. For example, G\"{o}bel et al., who further explored the effect of network delay on the SM strategy, demonstrated that SM will be more successful if every miner in the SM pool\footnotemark[3] helps propagate the hidden block \cite{GobelEtAl2016}. Kiayias et al. also showed that, under the game-theoretical setting, every miner will follow Nakamoto's mining protocol if no one has a hash rate greater than 30.8\% of the total hash rate \cite{KiayiasEtAl2016}.
% Moreover this type of strategic mining is inherent only in Proof-of-Work blockchain due to its probabilistic nature, and consequently other types of blockchain do not suffer from it \cite{ChenEtAl2017,KiayiasEtAl2017}.
\footnotetext[3]{A pool is a group of miners whose mining processes are coordinated such that they receive their individual mining rewards in a smaller chunk more frequently comparing to solo mining \cite{BitcoinWiki_Pool_2018}.}

On the contrary, a number of improvements of Nakamoto's mining protocol have been suggested, but they are difficult to implement in practice \cite{EyalSirer2014,ZhangPreneel2017}. In particular, the improvements which raises the hash rate required for SM to be effective needs a precise coordination among miners to adopt them at the same time.

Despite the significant body of work on SM, the idea of multiple miners individually and simultaneously employing SM has not been fully explored in the existing literature. In particular, most works so far studied SM or similar strategies in a setting with one malicious miner and one non-malicious miner \cite{EyalSirer2014,GervaisEtAl2016,KiayiasEtAl2016,NayakEtAl2016,SapirshteinEtAl2017,ZhangPreneel2017}. To our knowledge, there is a small-scale study which was recently conducted in parallel \cite{LiuEtAl2018}. Compared to their work, our findings are more robust due to a large number of malicious miners in the system and a fair treatment of the underlying network in our experiment. Our work also offers a game-theoretical analysis which is a natural extension when malicious miners are considered self-interested agents that act strategically to maximise their mining rewards.

% Despite the significant body of work on SM, the idea of multiple miners individually and simultaneously employing SM has not been explored in the existing literature. In particular, current work has only studied SM or similar strategies in settings with one malicious miner and one non-malicious miner \cite{EyalSirer2014,GervaisEtAl2016,KiayiasEtAl2016,NayakEtAl2016,SapirshteinEtAl2017,ZhangPreneel2017}. Since there are many miners in actual systems, it is important to investigate how SM will perform in a context of multiple miners that could individually employ SM and earn a higher mining reward than they should.

%%%%%%%%%%%%%%%%%%%%%%%%%%%%%%%%%%%%%%%%%%%%%%%%%%
\section{Models of the PoW Blockchain Mining}
\label{sec:model}
%%%%%%%%%%%%%%%%%%%%%%%%%%%%%%%%%%%%%%%%%%%%%%%%%%
In this section, we formally define two models of Proof-of-Work (PoW) blockchain mining where the difference between them lies in the miner's capability of choosing a mining strategy. To clearly observe the effect of varying number of miners upon SM, assumptions are made as follows:
\begin{enumerate}
 \item A fully connected network of miners without any communication delay;
 \item An equal amount of mining reward per block to the creator of every block in the blockchain; and
 \item The same computational difficulty (the target hash value) for every block in the blockchain. 
\end{enumerate}

We consider two mining strategies: \textit{Honest Mining (HM)} and \textit{Selfish Mining (SM)}\footnotemark[4]. The first is Nakamoto's mining protocol where a miner always mines and publishes a new block from the last block of the longest blockchain. On the other hand, the latter is a strategy that hides its recently created block to privately mine from it and then strategically publishes its hidden blocks to overwrite the others' blocks in the currently longest chain \cite{EyalSirer2014}. That is, whenever SM receives a new block created by the others, SM also publishes its block with the expectation that it reaches the rest of the network more quickly than the received block. If SM's block reaches first, the another block will be ignored. In addition, SM publishes all hidden blocks to completely overwrite the other chain whenever the SM's branch is longer by one. At any time, SM always mines the deepest block, and it abandons its chain entirely whenever another chain is longer.
\footnotetext[4]{We do not use the optimised (two-miners) SM \cite{GervaisEtAl2016,SapirshteinEtAl2017} since it might not be optimal in our context of multiple miners. The method of obtaining an optimal strategy in this context is also not yet known and lies outside the scope of this work.}

Subsequently there are three types of miners: Honest miner, Selfish miner, and Strategic miner. By definition, Honest miner and Selfish miner perform HM and SM respectively; henceforth HM and SM will also be used to denote them. In contrast, \textit{Strategic miner (StrM)} is a miner that uses either the HM or SM strategy depending on which maximises its mining reward. Note that StrM will be referred to only in the second model where we consider miners are capable of choosing their strategies.

%%%%%%%%%%%%%%%%%%%%%%%%%%%%%%%%%%%%%%%%%%
\subsection{Fixed Strategy Mining Model}
\label{subsec:fixedStrategyModel}
%%%%%%%%%%%%%%%%%%%%%%%%%%%%%%%%%%%%%%%%%5
Here, we describe the first model of the PoW blockchain mining process where every miner employs a fixed mining strategy. Formally, a Markov model of the fixed strategy mining $\mathcal{M} = \left( I, C, P, S, \mathbbm{P}\left(\cdot\right), \mathbbm{U}\left(\cdot\right) \right)$ is as follows:

\begin{itemize}
 \item $I = \left\lbrace 1,2,...,N \right\rbrace$ denotes a set of all miners individually represented by a positive integer.
 \item $C = \left( c_i | c_i \in \left\lbrace \text{HM},\text{SM} \right\rbrace , i \in I \right)$ is a list of miner's mining strategies where the $i$-th element is a mining strategy used by the $i$-th miner in $I$.
 \item $P = \left( p_i | p_i \in \left[ 0,1 \right], \sum_{i \in I} p_i = 1 , i \in I \right)$ is a tuple of miner's \textit{mining powers} where the $i$-th element is the $i$-th miner's proportion of the total hash rate. That is, $P$ is a power allocation or power distribution of miners in the system.
 \item $S$ is a set of all states in this Markov model where each element $s \in S$ is a state of the blockchain. Note that the initial state $s_0 \in S$ is the blockchain with only one block that is not owned by any miner in $I$.
 \item $\mathbbm{P}\left(\cdot\right)$ is a state transition function where its probability mass is $\mathbbm{P} \left( s_{t+1} | s_t \right) = p_i$, and the next state $s_{t+1}$ is the current state $s_t$ that includes the new block created by miner $i$. In other words, a transition from state $s_t$ to state $s_{t+1}$ represents a discovery of a new block with respect to miner's mining powers.
 \item $\mathbbm{U}\left(\cdot\right)$ is a utility function that gives the converged value of the proportion of a miner's blocks in the longest blockchain. Given a sufficiently long time $t$ and a state $s_t \in S$ that has only one longest chain of blocks, the $i$-th miner's \textit{mining reward} $\mathbbm{U} \left( s_t,i \right)$ can be computed as follows:
 \begin{equation}
  \label{eq:utility_function}
  \mathbbm{U} \left( s_t,i \right) = \frac{b_i}{\sum_{i \in I} b_i}
 \end{equation}
 where $b_i$ is the total number of $i$-th miner's blocks in the longest chain starting from the initial block in $s_0$. Since this is a stationary Markov model, there always exists the convergence time $t$ where $\forall t_1, t_2 \in \left[ t, \infty \right) : \left| \mathbbm{U} \left( s_{t_1},i \right) - \mathbbm{U} \left( s_{t_2},i \right) \right| \leq \alpha$ and $\alpha$ is a negligible positive number.
\end{itemize}

%%%%%%%%%%%%%%%%%%%%%%%%%%%%%%%%%%%%%%%%%%%%%%%%
\subsection{Dynamic Strategy Mining Model}
\label{subsec:dynamicStrategyModel}
%%%%%%%%%%%%%%%%%%%%%%%%%%%%%%%%%%%%%%%%%%%%%
In contrast to the previous model, the model here considers the malicious miner's capability of choosing a mining strategy that maximises their individual mining reward. With a game analysis of this model, we account for a change of the SM miner's strategy when they deem it is better off to use HM under some power allocations.

In particular, we extended the previous model such that every SM miner becomes a StrM miner who chooses their mining strategy given information of other miners' available strategies and all possible mining rewards that the StrM miner will receive. In particular, an empirical normal-form game of the PoW mining is denoted by $\mathcal{G} = \left( I, C', P, \mathbbm{A}\left(\cdot\right), \mathbbm{U}'\left(\cdot\right) \right)$ where $I$ and $P$ are the same as before and the rest are described as follows:
\begin{itemize}
 \item $C' = \left( c'_i | c'_i \in \left\lbrace \text{HM},\text{StrM} \right\rbrace , i \in I \right)$ is a list of miner's types where the $i$-th element indicates whether an $i$-th miner is a HM miner or a StrM miner.
 \item $\mathbbm{A}\left(\cdot\right)$ is a function that maps a type of miner to a set of strategies. Given an $i$-th miner's type $c'_i \in C'$, the function $\mathbbm{A}\left(\cdot\right)$ is formally described as follows. 
 \begin{equation*}
  \mathbbm{A}\left(c'_i\right) =
  \begin{cases}
  \left\lbrace \text{HM}, \text{SM} \right\rbrace & \quad \text{if } c'_i = \text{ StrM} \\
  \left\lbrace \text{HM} \right\rbrace & \quad \text{otherwise }
  \end{cases}
 \end{equation*}
 Consequently, a strategy profile is denoted as $A = \left( a_i | a_i \in \mathbbm{A}\left(c_i\right), c'_i \in C, i \in I \right)$ or $A = \left( a_i, a_{-i} \right)$ where $a_i$ is the $i$-th miner's strategy and $a_{-i}$ collectively denotes the rest.
 \item $\mathbbm{U}': I \times A^N \mapsto \left[ 0,1 \right]$  is a payoff function that computes a miner's mining reward. Given a strategy profile $A$, the value of $\mathbbm{U}' \left( i, A \right)$ is simply retrieved from the utility function $\mathbbm{U}$ of the previously described model $\mathcal{M}$ where its strategy list $C$ corresponds to $A \in \mathcal{G}$ and other elements of the model are the same.
\end{itemize}
%%%%%%%%%%%%%%%%%%%%%%%%%%%%%%%%%%%%%%%%%%%%%%%%%%
\section{Power Threshold, Safety Level and Equilibrium}
\label{sec:psneThresholdSafety}
%%%%%%%%%%%%%%%%%%%%%%%%%%%%%%%%%%%%%%%%%%%%%%%%%%
As mentioned in Section \ref{sec:introduction}, we are interested in the minimum mining power that enables SM/StrM miners to earn an unfairly large amount of mining reward and the minimum total sum of mining power of all HM miners that can prevent such an unfair outcome.

In more detail, an unfairly large mining reward in our models is one that exceeds the miner's mining power. Originally, a system of all HM miners, in the long run, will allocate a mining reward equal to a miner's mining power (since everyone mines from the latest block and the expected proportion of miner's blocks is the miner's power.) However, a miner with sufficiently high mining power can use SM and gain a mining reward that is higher than their mining power. Such an unfairly large reward is demonstrated in the next section of this paper.

In our discussion, we then look for a \textit{power threshold} which is the least mining power that lets SM/StrM earn its unfairly large mining reward regardless of how much mining power the others possess. Consequently, a SM/StrM miner whose mining power reaches the power threshold will always earn a mining reward that is more than they should.
\begin{mydef}\label{def:powerThreshold}
 Given $\hat{P}, \left( p \right)$ the set of all possible power allocations where a SM/StrM miner has mining power $p$, and $\mathbbm{U}_{p,P}$, the mining reward of the SM/StrM miner with mining power $p$ in a power allocation $P$, a \textbf{power threshold} $\beta$ is one that satisfies the following condition:
\end{mydef}
\begin{equation*}
 \beta = \min \left\{\right. p \ | \ \forall P \in \hat{P} \left( p \right) : \mathbbm{U}_{p,P} > p \ ; \ \ \forall q \in \left[ p , 1 \right], \forall P' \in \hat{P}\left(q\right) : \mathbbm{U}_{q,P'} > q \left.\right\}
\end{equation*}
In other words, for every SM/StrM's mining power that yields mining reward larger than the power regardless of the others' power, a power threshold is the least power of SM/StrM that also yields such a reward for every SM/StrM's power beyond the threshold.

Similarly, we also search for a \textit{safety level}, which is the least mining power of a collective of all HM miners that prevents all SM/StrM miners from earning their unfairly large mining reward. Once the safety level is reached, no SM/StrM miner will be able to gain a mining reward that is higher than their mining power.

\begin{mydef}
 Given $I'$, the set of all SM/StrM miners, and $\mathbbm{U}_{i, p_i}$, the mining reward of the $i$-th miner with mining power $p_i$, a \textbf{safety level} $\gamma$ is one that satisfies the condition below:
 \begin{equation*}
 \begin{split}
  \gamma = \min \left\{\right. p_{\text{HM}} \ | \ &\forall P \in \hat{P} \left( p_{\text{HM}} \right) , \forall i \in I' \left( P \right) : \mathbbm{U}_{i, P} \leq p_{i,P} \ ; \\
  &\forall q_{\text{HM}} \in \left[ p_{\text{HM}},1 \right] , \forall P' \in \hat{P} \left( q_{\text{HM}} \right) , \forall i' \in I' \left( P' \right) : \mathbbm{U}_{i', P'} \leq p_{i',P'} \left.\right\} 
 \end{split}
 \end{equation*}
 where $\hat{P} \left( p \right)$ is the same as in Definition \ref{def:powerThreshold} and $p_{\text{HM}}$ is a mining power sum of all HM miners.
\end{mydef}
That is, for every mining power of an HM collective that results in all SM/StrM's mining rewards being no greater than their power regardless of how much power all SM/StrM individually have, a safety level is the collective's least power that also yields such SM/StrM's rewards for every HM collective's power beyond the safety level.

In the dynamic strategy mining model (Section \ref{subsec:dynamicStrategyModel}), we will retrieve an outcome of the game prior to an analysis of the safety level and the power threshold. In particular, we use the concept of pure-strategy $\epsilon$-equilibrium ($\epsilon$-PE) to derive the choice of miners' strategies that maximises their mining reward. The concept is also useful to disregard small fluctuations in the payoff value; such a fluctuation is caused by a stochastic nature of the PoW blockchain mining process and consequently could lead us to misinterpret the result. 
\begin{mydef}\label{def:epsilonPE}
 A pure-strategy $\epsilon$-equilibrium ($\epsilon$-PE) where $\epsilon > 0$ is a strategy profile $A^{*} = \left( a^*_i, a_{-i} \right)$ that satisfies the following condition:
\end{mydef}
\begin{equation*}
 \forall i \in I , \forall a_i \in \mathbbm{A}\left(c'_i\right) : \quad \mathbbm{U}' \left( i, A^{*} \right) \geq \mathbbm{U}' \left( i, \left( a_i, a_{-i} \right) \right) - \epsilon
\end{equation*}
In other words, for each and every miner, there are no other mining strategies that allow them to gain a higher utility than the strategy in the pure-strategy $\epsilon$-equilibrium by $\epsilon$, given that the others' strategies are fixed.

Finally, an extra assumption where HM is more preferable to SM will be incorporated in the $\epsilon$-PE analysis of the result. In the next section, we show the existence of multiple equilibria due to a negligible difference between HM's and SM's mining reward in the same power allocation. Since there is neither an incentive nor a proper reason for StrM to use SM instead of HM in such cases, we disregard such equilibria with SM by the \textit{HM-preference assumption}, which is defined as follows:
\begin{mydef}\label{def:HM-pref}
 Given a pair of $\epsilon$-equilibria $A^{*} = \left( a^*_i , a_{-i} \right)$ and $A^{**} = \left( a^{**}_i , a_{-i} \right)$ where $a^*_i \neq a^{**}_i$ (one $i$-th miner's choice is HM and the another is SM) under the same instance of model $\mathcal{G}$, an HM-preferable $\epsilon$-equilibrium is the equilibrium where the $i$-th miner's choice is HM.
\end{mydef}

% \clearpage
%%%%%%%%%%%%%%%%%%%%%%%%%%%%%%%%%%%%%%%%%%%%%%%%%%
\section{Empirical Results and Discussion}
\label{sec:result}
%%%%%%%%%%%%%%%%%%%%%%%%%%%%%%%%%%%%%%%%%%%%%%%%%%
To address our research question, we carry out discrete event simulations of the models such that different numbers of SM/StrM miners and different power allocations are accounted.\footnotemark[5] Each simulation setting is also repeatedly simulated 100 times to compute an average of the converged utility value. In a rare case of non-convergence, we use the value at the 200,000th timestep, which is analogous to 3-4 years in the Bitcoin system and well approximates the system behaviour compared to the results of others \cite{EyalSirer2014,SapirshteinEtAl2017}.
\footnotetext[5]{Note that modelling the underlying network is out of scope of this work. Consequently multiple broadcast messages that occurred in single timestep were processed in a uniformly random manner.}
\begin{table}
 \centering
 \caption{Simulation parameters}
 \begin{tabular}{@{}ccccc@{}}
  \toprule
  \multicolumn{4}{c}{\textbf{Parameter}} & \multicolumn{1}{c}{\textbf{\ Value \ }} \\ \midrule
  \multicolumn{4}{c}{$\alpha$ \ (Equation \ref{eq:utility_function})} & \ 0.0001 \ \\
  \multicolumn{4}{c}{$\epsilon$ \ (Definition \ref{def:epsilonPE})} & \ 0.0001 \ \\
  \multirow{4}{*}{\ Power step \ } & for & 1,2,3 & SM/StrM cases & 0.01 \\
   & for & 4 & SM/StrM case & 0.02 \\
   & for & 5,6,7 & SM/StrM cases & 0.04 \\
   & for & 8,9 & SM/StrM cases & 0.05 \\
  \bottomrule
 \end{tabular}
 \label{tab:parameters}
\end{table}

Due to the extremely large number of required simulations, we carry out simulations only for the base parameters and perform permutation to cover all necessary results. To exemplify this, we swap the miner's utility values of the model $\mathcal{M}_1$ with $C_1 = \left( \text{HM},\text{SM} \right)$ and $P_1 = \left( 0.4,0.6 \right)$ and use it as a result of the model $\mathcal{M}_2$ where $C_2 = \left( \text{SM},\text{HM} \right)$ and $P_2 = \left( 0.6,0.4 \right)$. We also treated a collective of HM miners as a single HM miner since their individual earning is unnecessary in this work and an overall outcome of their individual mining is the same as mining done by one HM with their combined mining powers in our models.

%%%%%%%%%%%%%%%%%%%%%%%%%%%%%%%%%%%%%%%%%%%%%%%%%%
\subsection{Fixed Strategy Mining}
\label{subsec:fixStrategyMining}
\begin{figure}
 \centering
 \begin{subfigure}{0.49\linewidth}
  \centering
  \includegraphics[scale=1.1]{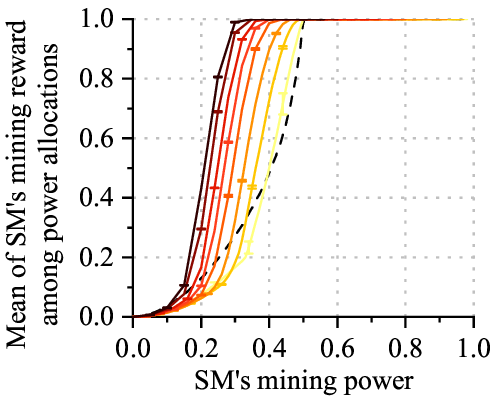}
  \vspace*{-30pt}
  \caption{}
  \label{fig:S3_smReward}
 \end{subfigure}
 \begin{subfigure}{0.49\linewidth}
  \centering
  \includegraphics[scale=1.1]{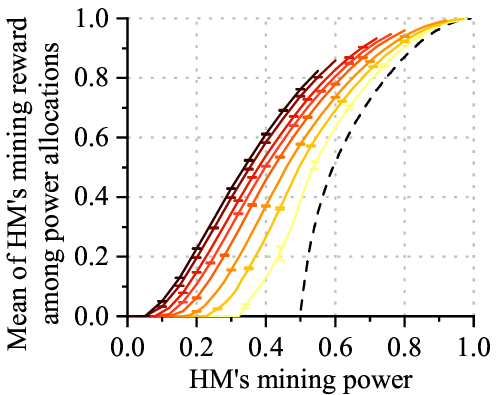}
  \caption{}
  \label{fig:S3_hmReward}
 \end{subfigure}
 \begin{subfigure}{\linewidth}
  \centering
  \vspace*{5pt}
  \includegraphics[scale=1.1]{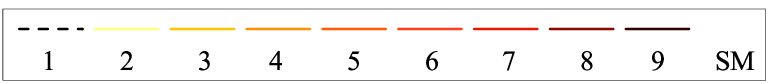}
 \end{subfigure}
 \caption{Average of the SM's mining reward among different power allocations with specific SM's mining power \subref{fig:S3_smReward} and an average of the HM's mining reward among different power allocations with specific HM's mining power \subref{fig:S3_hmReward} in a system with different numbers of SM. Standard error of the mean is shown as an error bar.}
 \label{fig:S3}
\end{figure}

%%%%%%%%%%%%%%%%%%%%%%%%%%%%%%%%%%%%%%%%%%%%%%%%%%
In general, the mining powers of SM and HM that yield an unfairly large mining reward decreases with the number of SM miners in the system. As shown in Figure \ref{fig:S3_smReward}, the mean of SM's mining reward among different power allocations exponentially grows in an increase of SM's mining power until its convergence at one. However, the range of SM's mining power during the exponential growth gradually decreases with the number of SM miners. A similar trend in the HM's mining reward with respect to the HM's mining power is also observed and shown in Figure \ref{fig:S3_hmReward}.

As shown by Liu et al. \cite{LiuEtAl2018}, our observation has also revealed a similar underlying cause of the trend of HM/SM's mining reward. Generally, the higher the number of miners, the less mining power each of them has. With a low mining power, SM is less likely to create a private chain longer than the other chains and therefore most of their computational resources are wasted. In turn, a mining power that HM/SM requires to earn an unfairly large mining reward become less in a system with a large number of miners. 

\begin{figure}
 \centering
 \includegraphics[scale=1.1]{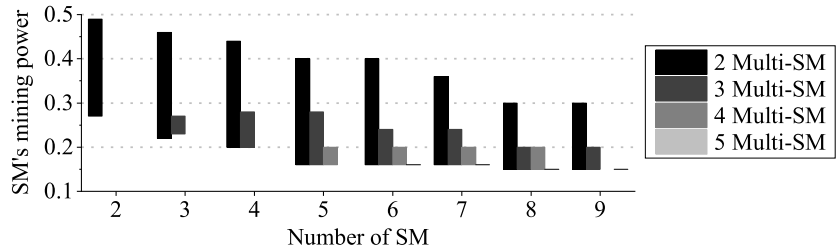}
 \caption{Ranges of mining power of multiple and profitable SM (Multi-SM) with respect to an increasing number of SM. The number of Multi-SM indicates a number of SM miners who simultaneously gain a mining reward that is more than their mining power. In such ranges, the individual mining powers of the SM are equal.}
 \label{fig:S3_multiSM}
\end{figure}

Surprisingly, a number of SM miners can simultaneously get their unfairly large rewards under some specific power allocations. In particular, for each number of SM miners, their mining powers in such a power allocation are the same and also larger than a certain value. However, the range of such mining powers decreases and shortens as the number of SM increases as shown in Figure \ref{fig:S3_multiSM}. We therefore hypothesise that this behaviour does not exist in a system with an extremely large number of miners.

%%%%%%%%%%%%%%%%%%%%%%%%%%%%%%%%%%%%%%%%%%%%%%%%%%
\subsection{Dynamic Strategy Mining}
\label{subsec:dynamicStrategyMining}
%%%%%%%%%%%%%%%%%%%%%%%%%%%%%%%%%%%%%%%%%%%%%%%%%%
In the previous section, we have shown that a SM miner with low mining power earns less than their power. However, they might be able to earn more if they switch back to HM instead. Such a switch can induce further strategy switches due to a change in mining reward. In this section, we use a game-theoretical concept of equilibria (as described in Section \ref{sec:psneThresholdSafety}) to tackle the strategy change and discuss the outcome.

\begin{figure*}
 \centering
 \begin{subfigure}{0.49\linewidth}
  \centering
  \includegraphics[scale=1.1]{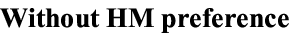}
  \vspace*{7pt}
 \end{subfigure}
 \begin{subfigure}{0.49\linewidth}
  \centering
  \includegraphics[scale=1.1]{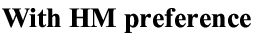}
  \vspace*{7pt}
 \end{subfigure}
 \begin{subfigure}{0.49\linewidth}
  \centering
  \includegraphics[scale=1.1]{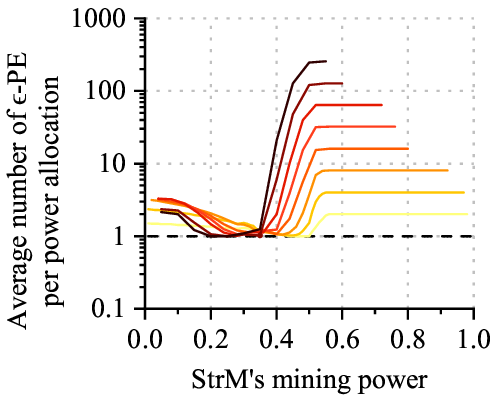}
  \caption{}
  \label{fig:S1A1_psnePerAllo}
 \end{subfigure}
 \begin{subfigure}{0.49\linewidth}
  \centering
  \includegraphics[scale=1.1]{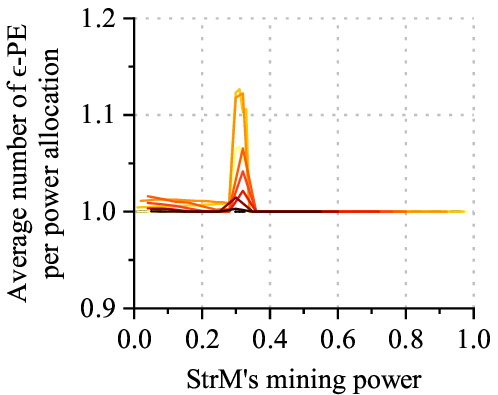}
  \caption{}
  \label{fig:S1A2_psnePerAllo}
 \end{subfigure}
 \begin{subfigure}{0.49\linewidth}
  \centering
  \includegraphics[scale=1.1]{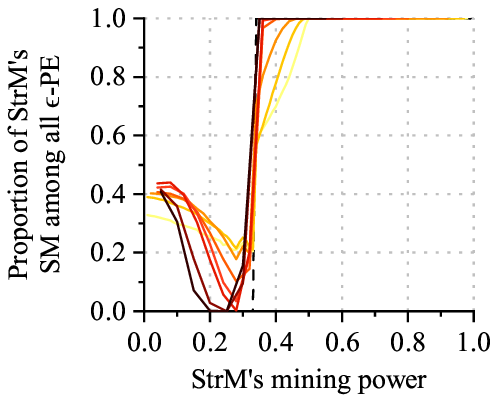}
  \caption{}
  \label{fig:S1A1_strategy}
 \end{subfigure}
 \begin{subfigure}{0.49\linewidth}
  \centering
  \includegraphics[scale=1.1]{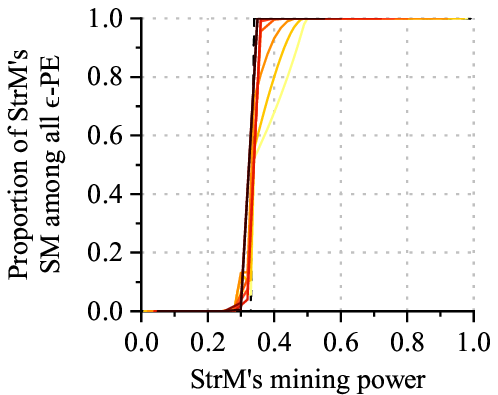}
  \caption{}
  \label{fig:S1A2_strategy}
 \end{subfigure}\\
 \medskip
 \begin{subfigure}{\linewidth}
  \centering
  \includegraphics[scale=1.1]{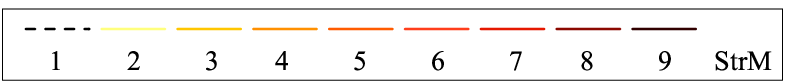}
 \end{subfigure}
 \caption{Average number of $\epsilon$-PE per power allocation (a,b) and an overall StrM's strategy in $\epsilon$-PE (c,d) with different number of StrM in the system. Left figures (a,c) are results not under the HM-perference assumption, while right figures (b,d) are results under the HM-preference assumption.}
 \label{fig:S1}
\end{figure*}

We first notice multiple equilibria for particular power allocations and at least one equilibrium for every power allocation in every number of StrM in the system. As shown in Figure \ref{fig:S1A1_psnePerAllo}, the average number of $\epsilon$-PE per power allocation is always at least one. However, it becomes extremely large in power allocations where there is a StrM with a relatively high mining power.

We observe that the large number of $\epsilon$-PE is caused by a StrM miner expressing an indifference between HM and SM strategy where there is another StrM with a particularly large mining power. In such a situation, there is no significant difference of mining reward between HM and SM used by the StrM with low mining power; which results in a moderate amount of StrM's SM over all $\epsilon$-PE as depicted in Figure \ref{fig:S1A1_strategy}. Consequently, the number of $\epsilon$-PE will simply be a combinatorial number of the StrM's HM/SM with low mining power and therefore grows in an increase of the number of StrM as shown in Figure \ref{fig:S1A1_psnePerAllo}.

With the HM-preference assumption, a reasonable choice of StrM's strategy in $\epsilon$-PE is obtained. In particular, StrM will no longer choose SM if there is no significant difference between HM's and SM's mining reward. The change of strategy in $\epsilon$-PE is clearly demonstrated by a comparatively low average number of PSNE per power allocation in Figure \ref{fig:S1A2_psnePerAllo} and no SM strategy chosen by StrM with mining power under 0.3 in Figure \ref{fig:S1A2_strategy}.

\begin{figure}
 \centering
 \begin{subfigure}{0.49\linewidth}
  \centering
  \includegraphics[scale=1.1]{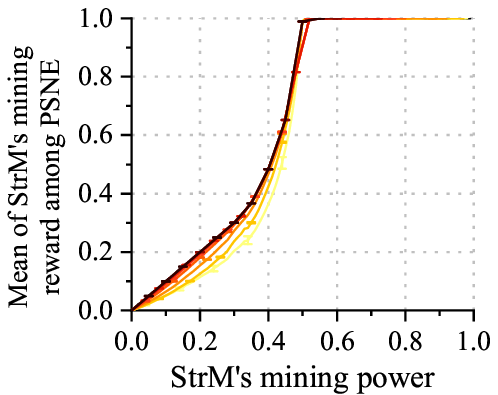}
  \caption{}
  \label{fig:S1A2_strMReward}
 \end{subfigure}
 \begin{subfigure}{0.49\linewidth}
  \centering
  \includegraphics[scale=1.1]{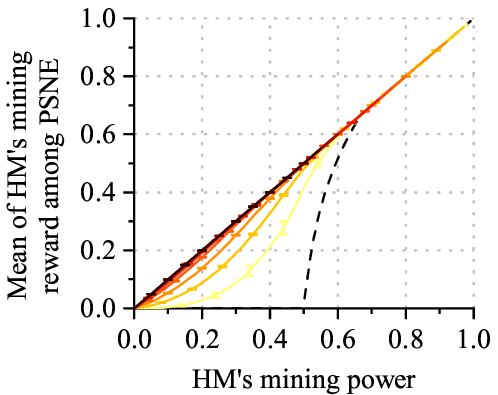}
  \caption{}
  \label{fig:S1A2_hmReward}
 \end{subfigure}
 \begin{subfigure}{\linewidth}
  \centering
  \vspace*{7pt}
  \includegraphics[scale=1.1]{figure/Legend1}
 \end{subfigure}
 \caption{Average of the StrM's mining reward among different $\epsilon$-PE with specific StrM's mining power \subref{fig:S1A2_strMReward} and average of the HM's mining reward among different $\epsilon$-PE with specific HM's mining power \subref{fig:S1A2_hmReward} in a system with different numbers of StrM. Standard error of the mean is shown as an error bar.}
 \label{fig:S1A2_reward}
\end{figure}

Clearly, StrM choose SM more than HM as their mining power increases to enjoy a larger amount of mining reward. This speculation is confirmed in Figures \ref{fig:S1A2_strategy} and \ref{fig:S1A2_strMReward}. That is, StrM starts to choose SM more once their mining power exceeds one-fourth. Once StrM possesses at least half of the total mining power, they always choose SM to earn the whole mining reward from the system.

Interestingly, the more StrM in the system, the more their choice of mining strategy and their mining reward becomes similar to the case of single StrM. As demonstrated in Figure \ref{fig:S1A2_strategy}, when the number of StrM miners increases, the transition of the StrM's strategy from HM to SM gradually becomes sharper similarly to the case of single StrM. Likewise, Figure \ref{fig:S1A2_strMReward} shows a convergence of the mining reward of StrM with mining power lower than \sfrac{1}{2} to one of the case of single StrM.

In contrast, the HM's mining reward does not converge to one in the case of single StrM. Instead, it converges to their mining power as the number of StrM increases. This is empirically shown in Figure \ref{fig:S1A2_hmReward}, where the mining reward of HM's mining power under 0.67 asymptotically approaches their mining power as the number of StrM increases.

Even with the HM-preference assumption, there still are multiple $\epsilon$-PE for particular power allocations. Such multiple equilibria are shown in Figure \ref{fig:S1A2_psnePerAllo} where an average number of $\epsilon$-PE per power allocation is more than one for any StrM's mining power ranging up to 0.36. We found that multiple StrM with the same mining power larger than 0.3 together choose either HM or SM in such $\epsilon$-PE. Since an individual deviation from HM to SM or vice versa yields a comparatively low mining reward for the deviating StrM, multiple $\epsilon$-PE with such StrM together choosing HM or SM are formed.

\begin{figure}
 \centering
 \includegraphics[scale=1.1]{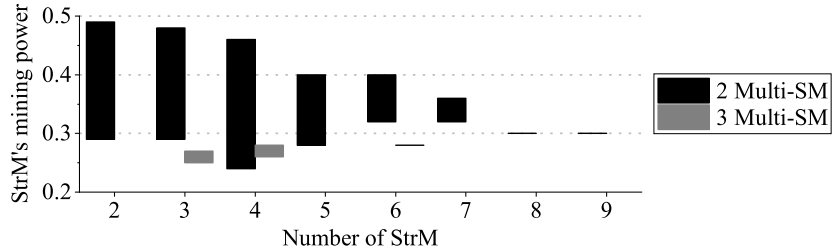}
 \caption{Ranges of mining power of multiple and profitable SM (Multi-SM) with respect to an increasing number of StrM. The number of Multi-SM indicates a number of StrM miners who individually use SM and simultaneously gain mining reward more than their mining power. In such ranges, the individual mining powers of such StrM are equal.}
 \label{fig:S1A2_multiSM}
\end{figure}

On further inspection, the $\epsilon$-PE where multiple SM are chosen by StrM becomes less likely to occur as the number of StrM increases. Compared to the fixed strategy model's, the range of mining power of multiple SM in this model is even less. As shown in Figure \ref{fig:S1A2_multiSM}, a mining-power range of multiple StrM miners that possess nearly equal power and together choose SM in $\epsilon$-PE shortens in an increase of the number of StrM. Therefore, it is clear that this multiple SM is highly unlikely to occur in the presence of large number of StrM.

%%%%%%%%%%%%%%%%%%%%%%%%%%%%%%%%%%%%%%%%%%%%%%%%%%
\subsection{Safety Level and Power Threshold}
\label{subsec:threshSafety}
\begin{figure}
 \centering
 \includegraphics[scale=1.1]{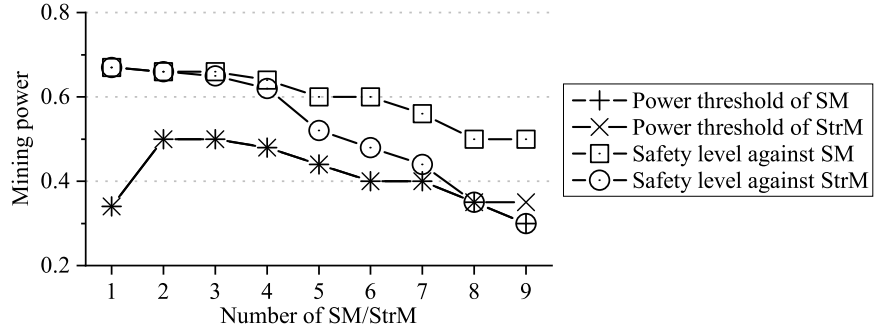}
 \caption{Power thresholds and safety levels with respect to different numbers of SM/StrM in the system. No difference of a safety level between one with the HM-preference assumption and one without the assumption is found.}
 \label{fig:thresSafety}
\end{figure}

%%%%%%%%%%%%%%%%%%%%%%%%%%%%%%%%%%%%%%%%%%%%%%%%%%
As shown in Figure \ref{fig:thresSafety}, the safety level against SM/StrM monotonically decreases as the number of SM/StrM grows. Since the mining power that one miner possesses will decrease in an increasing number of miners in the system, SM/StrM with low mining power will become prominent. Such SM/StrM are unable to frequently create a private chain longer than the others (Section \ref{subsec:fixStrategyMining}) and consequently choose HM to maximise their rewards (Section \ref{subsec:dynamicStrategyMining}). As a result, the total mining power of miners performing HM increases, and the HM miner requires less mining power to prevent SM/StrM.

Moreover, the safety level is upper bounded by the case of one SM/StrM; that is, it is no greater than \sfrac{2}{3}. Intuitively, the case of one SM/StrM is the most difficult to prevent since it is a coalition of all SM/StrM miners combining their mining power and working together against HM miners. The safety level in this case is therefore the greatest one.

Similarly, the power threshold of SM/StrM decreases in an increasing number of SM/StrM after the case of single SM/StrM in the system. Due to SM with low mining power constantly wasting their effort, the amount of mining power which is required to secretly build the longest chain becomes less in turn. 

However, the power threshold of StrM is strictly lower bounded at \sfrac{1}{3}. A similar rationale can be applied here: HM with a mining power of \sfrac{2}{3} is the most difficult for SM strategy and therefore a mining power of \sfrac{1}{3} is at least required. On the contrary, StrM with mining power lower than \sfrac{1}{3} will always choose HM, as shown in Figure \ref{fig:S1A2_strategy}.

Clearly an upper bound of the power threshold of SM/StrM is \sfrac{1}{2}, which corresponds to Nakamoto's analysis \cite{Nakamoto2008}. Any mining power beyond the threshold always allows SM/StrM to successfully create the longest chain.
%%%%%%%%%%%%%%%%%%%%%%%%%%%%%%%%%%%%%%%%%%%%%%%%%%
\section{Conclusions and Future Work}
\label{sec:conclusion}
%%%%%%%%%%%%%%%%%%%%%%%%%%%%%%%%%%%%%%%%%%%%%%%%%%
In this work, an empirical investigation of the Selfish Mining (SM) strategy employed by multiple miners has been carried out. We separately considered two types of malicious miners where one (SM miner) always follow SM and the another (StrM miner) chooses to follow either Nakamoto's mining protocol or the SM strategy depending on which maximises its mining reward. Since our work accounted for multiple miners and a large number of malicious miners in the system, our findings (such as the case of multiple miners simultaneously and individually performing SM) are more practical than the other's so far. 

The effectiveness of the SM strategy varies when different types and different numbers of malicious miners are considered. In general, SM is more effective in the presence of a large number of SM miners since it can reap a larger amount of mining reward with the same hash rate in a system with StrM miners. However, SM in a system with a low number of StrM miners is less effective than one in a system with SM miners since it yields a smaller mining reward with the same hash rate.

Regardless of the type and the number of miners in the system, the least hash rate to perform SM and to prevent SM are no greater than \sfrac{1}{2} and \sfrac{2}{3} respectively. Additionally, both amounts of hash rate monotonically decrease in an increase of the number of malicious miners in the system. If only StrM miners are considered, then the least hash rate required for SM is strictly \sfrac{1}{3}. However, such an amount reduces further than \sfrac{1}{3} (as originally reported by Eyal and Sirer \cite{EyalSirer2014}) if SM miners are considered.

Despite the aforementioned, our result suggests that PoW blockchain systems are required to have a large number of miners to be more secure against SM. Since blockchain miners are working to earn their mining reward, they are utility-maximising agents or StrM miners in our model. As shown in Section \ref{subsec:dynamicStrategyMining}, SM is comparably less chosen in the presence of a large number of StrM miners. Together with the decreasing hash rate required for preventing SM, it can be concluded that a large number of miners can prevent SM and possibly similar malicious mining strategies.

A number of interesting questions still remain to be further investigated. As pointed out by Eyal and Sirer \cite{EyalSirer2014}, a network capability of SM miners is also an important factor that affects the effectiveness of SM. This aspect will be taken into account in our future work. Moreover, an optimal SM strategy in the context of multiple miners, similar to that in the work of Sapirshtein et al. \cite{SapirshteinEtAl2017}, is not yet known. With the optimal strategy, it remains to be seen whether our findings are still valid.
%%%%%%%%%%%%%%%%%%%%%%%%%%%%%%%%%%%
\subsubsection{Acknowledgement}
%%%%%%%%%%%%%%%%%%%%%%%%%%%%%%%%%%%
The authors gratefully acknowledge financial support from the EPSRC Doctoral Training Partnership, and the use of IRIDIS High Performance Computing Facility at the University of Southampton. We also would like to express our gratitude to all anonymous reviewers for their insightful comments.
\bibliographystyle{splncs04}
\bibliography{bibliography}
\end{document}